\documentclass[11pt]{article}

\usepackage{amsmath, amsfonts,
amssymb,latexsym,wasysym,graphicx}
\input epsf.sty

\newtheorem{Theorem}{Theorem}[section]
\newtheorem{Lemma}{Lemma}[section]

\newtheorem{Definition}[Lemma]{Definition}

\newcommand{\BEQ}{\begin{equation}}     % Gleichungen Anfang ..
\newcommand{\BEA}{\begin{eqnarray}}
\newcommand{\BD}{\begin{displaymath}}
\newcommand{\EEQ}{\end{equation}}       % .. und Ende
\newcommand{\EEA}{\end{eqnarray}}
\newcommand{\ED}{\end{displaymath}}
\newcommand{\del}{\delta}
\newcommand{\Del}{\Delta}
\newcommand{\eps}{\varepsilon}          % epsilon

\newcommand{\supp}{{\mathrm{supp}}}

\newcommand{\Vol}{{\mathrm{Vol}}}
\newcommand{\R}{\mathbb{R}}

\newcommand{\Z}{\mathbb{Z}}
\newcommand{\N}{\mathbb{N}}
\newcommand{\D}{\mathbb{D}}

\newcommand{\Id}{{\mathrm{Id}}}

\def\sgn{{\mathrm{sgn}}}

\newcommand{\eop}{\hfill $\Box$}        % quod erat demonstrandum ...

\newcommand{\II}{{\rm i}}               % gerades i fuer komplexe Einheit
\newcommand{\half}{{1\over 2}}          % 1/2 als Bruch
\renewcommand{\vec}[1]{\boldsymbol{#1}} % Vektoren fettgedruckt
\catcode`\@=11
\def\numberbysection{\@addtoreset{equation}{section}
        \def\theequation{\thesection.\arabic{equation}}}
\numberbysection

\begin{document}

\vspace*{1.5cm}
\begin{center}
{\Large \bf Minkowski curvelets and wave equations}

\end{center}

\vspace{2mm}
\begin{center}
{\bf  J\'er\'emie Unterberger}
\end{center}

\vspace{2mm}
\begin{quote}

\renewcommand{\baselinestretch}{1.0}
\footnotesize
{We define a new type of wavelet frame adapted to the study of wave equations, that we call {\em Minkowski curvelets}, by reference to the curvelets introduced by Cand\`es, Demanet and
Donoho. These space-time, strongly anisotropic, directional wavelets have a Fourier support which does not intersect the light-cone; their maximal size is proportional to the inverse
of the distance to the light-cone. We show that the matrix of the Green kernel of the Klein-Gordon operator on Minkowski space-time has a nearly exponential off-diagonal decay in this
basis.    
}
\end{quote}

\vspace{4mm}


\tableofcontents

%\medskip

%%%%%%%%%%%%%%%%%%%%%%%%%%%%%%%%%%

\section{Introduction}

%%%%%%%%%%%%%%%%%%%%%%%%%

Let $\Box$  denote a wave operator on $\R\times\R^d$, $d\ge 1$ of the form
\BEQ \Box=\partial_t^2-A(x,\nabla_x)\EEQ
where $A(x,\nabla_x)$ is a time-independent, uniformly elliptic partial differential operator of order $2$ on $\R^d$.

As well known, if $v:\R_+\times\R^d\to\R$ solves the wave equation with initial condition
\BEQ \Box v(t,x)=0,\qquad v(0,x)=0,\ \partial_t v(0,x)=F(x)\qquad (t\ge 0) \EEQ
where $F:\R\to\R^d$ is a smooth function rapidly decaying at infinity,
then, letting $v(t,x)=(E_1(t)F)(x)$,
 \BEQ w(t,x):=\int_{0}^t (E_1(t-s)F(s,\cdot))(x) ds  \label{eq:E_1*F}\EEQ
 solves the equation $\Box w(t,x)=F(t,x)$ on $\R\times\R^d$.  We write $w=G(F)$  and call $G$ the {\em Green kernel} of $\Box$. 
 
 \medskip

The problem we are interested in, and solve in the case of the Klein-Gordon operator on Minkowski space-time,  is the following. Can one find a $\Box$-independent  'optimal' Riesz basis  $(\psi_{\Del})_{\Del\in\D}$ of $L^2(\R\times\R^d)$ \cite{HerWei}, indexed by
some set $\D$,  in which the matrix
coefficients of the Green kernel,
\BEQ G_{\Del,\Del'}:=\langle \psi_{\Del},G\psi_{\Del'}\rangle, \qquad \Del,\Del'\in\D \EEQ 
are essentially diagonal, with fast off-diagonal decay ?

\medskip

The principal motivation for this work comes from the rigorous study of real-time quantum field theory \cite{Erice,Sal,Riv,FMRS,FMRT}. From the beginning, constructive methods in QFT have been applied to imaginary-time, i.e.
Wick-rotated theories, thereby replacing hyperbolic wave operators by elliptic operators whose study is much simpler. Alas, real physics lives in real-time and is founded on perturbations
of linear, hyperbolic operators: Schr\"odinger operators describing quantum many-body Hamiltonian systems), or wave-like operators 
describing relativistic particles of high-energy physics \cite{PesSch}. The relevant quantities, including the simplest of all, the Green kernel, are then  oscillatory integrals which cannot
be appropriately bounded pointwise but only {\em in average}. Thus this article gives a satisfactory representation of the Green operator for the Klein-Gordon operator on Minkowski
space-time, a good start, hopefully, for the constructive study
of quantum field theory on pseudo-Riemannian manifolds. We also hope that such bases eventually turn out to be useful in the mathematical study of nonlinear wave equations  (global
well-posedness, blow-up, global parametrices, see e.g. \cite{Sog} or \cite{Tat0} for an old review) or of dispersive estimates for linear wave equations with variable coefficients, such as the wave equation on curved space-time, or
on domains with a boundary \cite{BouTzv,Leb,MetTat}.

\medskip

A very similar problem has been studied by Emmanuel Cand\`es, Laurent Demanet and David Donoho \cite{CanDem,CanDon} a few years ago.   They were looking for an optimal representation of 
the solution operator $E(t):u_0(\cdot)\mapsto u(t,\cdot)$ of linear symmetric
systems of first-order, hyperbolic differential equations of the form 
\BEQ \left(\frac{\partial }{\partial t}+\sum_{k=1}^d A_k(x) \frac{\partial }{\partial x_k}+B(x)\right)u=0, \quad u(0,x)=u_0(x)\in\R^n. \EEQ
The Riesz basis they constructed is made up of generalized, highly {\em anisotropic} wavelets for which they coined a nice name, {\em curvelets} . The matrix representation of the solution operator is  not diagonal, but rather similar to a {\em permutation matrix},  curvelets being shifted in phase space according  to the
underlying semi-classical  Hamiltonian flow obtained by the geometric optics approximation. Cand\`es et al. wrote
a  series of papers starting from 2004 on the remarkable properties of these curvelets,  both theoretical studies (representation of Fourier integral operators \cite{CanDem2} and applications to the mathematical representation of images with edges; curvelets are now commonly used for algorithms of compression of numerical photographies, which are much more
efficient than those based on the conventional, isotropic wavelets. The interested reader may refer to the web-page of E. Cand\`es (http://www-stat.stanford.edu/\~\,candes/publications.html)  for more information and references. Curvelets may be
seen as a phase-space extension of the second Paley-Littlewood dyadic decomposition used by Fefferman \cite{Fef}, Seeger, Sogge and Stein \cite{SSS} to obtain $L^p$ estimates for Fourier
integral operators. More recent references are two papers by Hart F. Smith \cite{Smi1,Smi2} (followed by a series of papers by Tataru \cite{Tat1,Tat2,Tat3} relying on a totally
different method); in the first reference, Smith proved Strichartz estimates for wave operators with $C^{1,1}$  coefficients by introducing
a kind of curvelet decomposition which partially inspired the work by Cand\`es et al.

The main feature of curvelets is that they are {\em directional, anisotropic} objects with a {\em parabolic scaling}, namely, their 'size' is   $O(2^{-j})$ in one direction, equally distributed on
the unit sphere, and $O(2^{-j/2})$ in the $d-1$ transversal directions;  'size' refers to the bulk of the support of the curvelets since they are not compactly supported. Then matrix coefficients of the solution operator $E(t)$ have a fast decay off the shifted diagonal in terms of an appropriate {\em pseudo-distance} between the bulks of the supports.

\bigskip

Of course, scalar wave operators may be converted into $2\times 2$-systems of first-order, hyperbolic differential equations, and our operator $E_1(t)$ is essentially a particular case of $E(t)$.
However our problem differs, in that we are interested in the Green kernel, given by the time convolution by $E_1$ as in eq. (\ref{eq:E_1*F}). Whereas the {\em curvelet} basis that Cand\`es,
Demanet and Donoho
constructed are functions of space coordinates only, we define a basis made up of functions of {\em space-time}. Quite naturally -- in view of our application to wave operators --, we called
these new objects {\em Minkowski curvelets}. 

Minkowski curvelets have {\em three} scales instead of {\em two}, given in terms of {\em two independent scale indices} $m,j$ called respectively {\em main scale} and {\em secondary scale}. As in the case of curvelets, they are not compactly supported,
but they have a fast decrease outside a small space-time window corresponding to the  'bulk' of their support. The space-time windows have largest size $O(2^{-m+j/2})$  along light rays ({\em longitudinal direction}), corresponding to  quasi-classical propagation in phase space. Their size along the  independent {\em transversal}, spatial {\em directions} is $O(2^{-m})$. Finally, going away from the direction of propagation in an
arbitrary {\em off-propagation direction}, curvelets fade away after a distance $O(2^{-m-j/2})$. Restricting to spatial directions and rescaling by typical propagation   
distances $O(2^{-m+j/2})$, one obtains the same ratios $O(2^{-j}), O(2^{-j/2})$ as for curvelets.

\bigskip

Our main result is the following. We let $(\psi_{\Del})_{\Del\in\D}$ be a fixed basis of Minkowski curvelets. To each index $\Del$ are associated in particular two scale indices,
the main scale $m_{\Del}$ and the secondary scale $j_{\Del}$. We define in section 2 a scaled 'distance' $d(\Del,\Del')$ between
two curvelets with indices $\Del,\Del'$. 

\newpage

{\bf Main theorem.}

{\em Let $G$ be the Green kernel of the Klein-Gordon operator $\Box_{\mu}:=\partial_t^2-\sum_{i=1}^3 \partial_{x_i}^2-\mu^2$ on Minkowski space-time, with mass $\mu\ge 0$. 
Let $\Del\in\D^{m,j},\Del'\in\D^{m',j'}$, with $m,m'\in\Z$, $j,j'\ge 1$, $m+j/2,m'+j'/2\ge 0$. Then, for every $N\ge 0$, there exists a constant $C_N>0$ such that
\BEQ |\langle G\psi_{\Del},\psi_{\Del'}\rangle|\lesssim 2^{-2\sup(m,m')} d(\Del,\Del')^{-N}. \label{eq:main} \EEQ
This result holds true with a slightly different definition of the Minkowski curvelets, adapted to the Klein-Gordon operator.
}

\bigskip

Assuming
$m_{\Del}=m_{\Del'}=m$ and $j_{\Del}=j_{\Del'}=j$, and using the relativistic notation $x_0$ for $t$, Fourier variables $(\xi_i)_{i=0,\ldots,3}=(\xi_0,\vec{\xi})$ scale  like 
$||\xi_0|-|\vec{\xi}||\approx 2^{m-j/2},\ |\xi_0|+|\vec{\xi}|\approx 2^{m+j/2}$ on the support of $\hat{\psi}_{\Del},\hat{\psi}_{\Del'}$, so $\xi^2:=\xi_0^2-\sum_{i=1}^3 \xi_i^2$ (the symbol of $\Box_0$) scales like $2^{2m}$ and a rapid estimate yields
$\hat{G}\approx\frac{1}{\xi^2}\approx 2^{-2m}$, accounting for the scaling factor in eq. (\ref{eq:main}). The interesting point is that the matrix $G_{\Del,\Del'}$ has an almost
exponential off-diagonal decay. In other words, the Fourier cut-offs destroy the  propagation of the waves, at least in the ultra-violet region $m+j/2,m'+j'/2\ge 0$ where $|\vec{\xi}|$ is large.

\medskip Our
estimates in the theorem are optimal in a region $|\xi_0|,|\vec{\xi}|\gg 1,\ \frac{|\xi_0|}{|\vec{\xi}|}\approx 1$ where geometric optics is a valid approximation. They are not
optimal (or even wrong) in the irrelevant regions $|\xi_0|\gg |\vec{\xi}|\gg 1$ or $|\vec{\xi}|\gg |\xi_0|\gg 1$ corresponding to $j=0$, and in the infra-red region $m+j/2,m'+j'/2<0$ 
where $|\xi_0|$ and $|\xi|$ are not
large enough and terms of lower order in the wave operator become essential.  A  companion article in preparation is devoted to complete, optimal estimates of the Green kernel
of the Klein-Gordon operator $\Box_{\mu}$ on $\R\times\R^d$. It requires the introduction of yet another family of curvelets, {\em Schr\"odinger} or {\em parabolic curvelets}, for
the study of the infra-red region, corresponding to the bottom of the mass hyperboloid $|\xi_0|\sim \mu,|\vec{\xi}|\ll \mu$.

It is reasonable to hope that the same type of result (or maybe a weaker one, maybe with a restriction to a subset of scales) is true for more general wave operators, such as e.g. the Klein-Gordon operator $\partial_t^2-\Del-\mu^2$ on
$\R\times M$, where $M\simeq \R^d$ is equipped with a Riemannian metric which is asymptotically flat at infinity \cite{MetTat}. Preliminary computations show that a naive short-time expansion of
the solution map $E_1(t)$ using Fourier integral operators does not give precise enough estimates, even for bounded times, if $\Box$ is a general wave operator. Obviously, the precise
definition of the curvelets should depend on the operator itself, since this is already the case for the Klein-Gordon operator on flat space-time. Of course one expects that the basis of
curvelets does not need to be changed for 'weak' perturbations of a given wave operator. The difficulty is that a semi-classical analysis is not sufficient to get precise estimates.
\bigskip

Our construction relies heavily on previous work by Cand\`es, Demanet and Donoho, in particular on reference \cite{CanDem}.  The article is organized as follows. We define  our new objects, {\em Minkowski curvelets}, in section 1, extending
{\em en passant} the construction of usual curvelets to arbitrary dimension (originally they were introduced in dimension 2, essentially for simplicity of exposition). Properties of Minkowski curvelets,
in particular precise decay estimates,
are developed in section 2. Section 3 is devoted to the Main Theorem.
 
\bigskip

{\em Notations.} If $t\in\R$,  $\langle t\rangle:=1+|t|$ and $\langle\langle t\rangle\rangle:=1+|t|+|t|^{-1}$. $(\chi^j)_{j\in\Z}$ is a  smooth multi-scale Fourier partition of unity (see Appendix).
Minkowski signature is by convention $(+,-,-,-)$, i.e. $x\cdot \xi=x_0 \xi_0-x_1 \xi_1-x_2\xi_2-x_3\xi_3$ and $x^2=(x^0)^2-x_1^2-x_2^2-x_3^2$, $\xi^2=\xi_0^2-\xi_1^2-\xi_2^2-\xi_3^2$.
The spatial components of $x$, resp. $\xi$ are denoted by $\vec{x}$, resp. $\vec{\xi}$; the reader should be warned against mistaking $x^2$ (Minkowski squared norm) for $|\vec{x}|^2$
(Euclidean squared norm of spatial components), while $|x^2|$ is simply the absolute value of $x^2$. The Fourier transform is defined with respect to the Minkowski scalar product, namely, ${\cal F}f(\xi)=\hat{f}(\xi)=\int e^{\II \xi\cdot x} f(x)dx$. The symbol $x \lesssim y$ means: $|x|\le C|y|$ for some universal constant $C$. The symbol $x\approx y$ means: $C^{-1}|x|\le |y|\le C|x|$ for
some universal constant $C>1$. 

%%%%%%%%%%%%%%%%%%%%%%%%%%%%%%%%%%%%%%%%%%%%
%%%%%%%%%%%%%%%%%%%%%%%%%%%%%%%%%%%%%%%%%%%

%%%%%%%%%%%%%%%%%%%%%%%%%%%%%%%%%
%%%%%%%%%%%%%%%%%%%%%%%%%%%%%%%%

\section{Curvelets: a reminder}

%%%%%%%%%%%%%%%%%%%%%%%%%%%%%%%%%%%%%%%%%%%
%%%%%%%%%%%%%%%%%%%%%%%%%%%%%%%%%%%%%%%%%%%

Cand\`es-Donoho-Demanet \cite{CanDem,CanDon}, relying on estimates for parametrices of wave equations previously  developed by Hart F. Smith \cite{Smi1}, introduced a family of generalized wavelets on $\R^2$
which they called {\em curvelets}. As we mentioned in the Introduction, curvelets  are appropriate for the study of parametrices of  wave equations (i.e. for the solution of wave equations
with initial conditions), in that the matrices $e^{\II tA}:=(\langle e^{\II t\sum_k A_k(x)\partial_{x_k}}\psi_{\Del},\psi_{\Del'}\rangle)_{\Del,\Del'}$ are maximally sparse, i.e. 
$(e^{\II tA})_{\Del,\Del'}$ is maximal when $\Del'=\Del(t)$ is the image of $\Del$ by the underlying semi-classical Hamiltonian flow, and is fast decreasing in the distance between
$\Del'$ and $\Del(t)$ otherwise.

\subsection{About wavelets and frames}

By wavelets we mean {\em orthonormal wavelets}. The profane reader  may refer e.g. to \cite{HerWei,Pin} for a simple and pedagogical introduction to real harmonic analysis and wavelets, and
look up in the classical references \cite{Mey,Ste} for detailed proofs and more advanced stuff. Let us simply recall here the following definitions and properties of wavelets.

\medskip

A (one-dimensional) {\em wavelet} is a function  $\psi\in L^2(\R)$   such that the functions $(\psi^{j,k})_{j\in\Z,k\in\Z}$ defined by Fourier rescaling and scaled translations in real space,
$\psi^{j,k}(x):=2^{j/2}\psi(2^{j}(x-k2^{-j}))$, make up an orthonormal basis of $L^2(\R)$. 

The  following two equations characterize wavelets (see \cite{HerWei}, chap. 7):
\BEQ \sum_{j\in\Z} |\hat{\psi}(2^{-j} \xi)|^2=1,\quad \xi\in\R^d \label{eq:basic-eq} \EEQ
\BEQ \sum_{j\in\Z} \hat{\psi}(2^{-j} \xi) \overline{\hat{\psi}(2^{-j} (\xi+2m\pi))}=0,\quad \xi\in\R,\quad m\in 2\Z+1.\EEQ

We shall only use eq. (\ref{eq:basic-eq}) that we shall call the {\em first basic  wavelet equation}.

\medskip

There are many different constructions of wavelets. The easiest ones are obtained by multiresolution analysis through Fourier analysis, and are compactly supported in Fourier space. Assume, say, that $|\supp\,  \hat{\psi}|\subset [1/2,2]$. Note that $\hat{\psi}^{j,k}(\xi)=2^{-j/2}e^{-\II k 2^{-j}\xi} \hat{\psi}(2^{-j}\xi)$; hence $|\supp\, \hat{\psi}^{j,k}|\subset [2^{j-1},2^{j+1}]$ independently
of the {\em space localization index} $k$; then $j$ plays the role of a {\em momentum} (or Fourier) {\em scale}. Lemari\'e and Meyer constructed a family of wavelets along these lines, depending
on the choice of a more or less arbitrary 'bell function' satisfying some support and symmetry properties, and satisfying the above support property. Provided the bell function is smooth,
$\hat{\psi}$ is compactly supported and smooth, hence $\psi$ is fast decreasing at infinity, i.e.   $|\psi^{j,k}(x)|\le C_r \langle
 x-k2^{-j}\rangle^{-r}$ for every $r\ge 0$, with a constant $C_r$ depending on $r$.  In other words, $\psi^{j,k}$ is centered around a window (an interval in fact), say, $[(k-1)2^{-j},(k+1)2^{-j}]$, of size $\approx 2^{-j}$, and has fast decrease outside
 this window.  Characterizing a wavelet $(\psi^{j,k})$ by its window $\Del^{j,k}:=[(k-1)2^{-j},(k+1)2^{-j}]$, one may index the orthonormal basis of wavelets by the windows,
 $(\psi^{j,k})_{j,k}=(\psi_{\Del})_{\Del\in \D}$, where $\D=\amalg_{j\in\Z}\D^j$ and $\D^j$ is the set of intervals of length $2^{1-j}$ centered at points which are integer multiples of $2^{-j}$. 
 This is the type of wavelets we shall be using. Later on we shall introduce generalized wavelets ({\em curvelets} or {\em Minkowski curvelets}) for which the couple of indices $(j,k)$ becomes
 some more complicated set of indices, and it will be more convenient to write the wavelet family as $(\psi_{\Del})_{\Del\in\D}$, where $\Del$ is a window which is some compact subset
 of $\R^d$, and $\Del$ ranges in some complicated set $\D$.
 
\medskip

For many purposes, it is not really necessary that $(\psi^{j,k})$ constitute an orthonormal basis. It is enough that they should make up a {\em tight frame}  (or {\em frame} for short
in the sequel). A frame $(f^j)_{j\in I}$, $I\simeq\Z$, is a family of elements of a Hilbert space $\cal H$ such that the Parseval identity is satisfied, i.e.
 $||\phi||^2=\sum_j \langle \phi,f^j\rangle^2$, or equivalently (see \cite{HerWei}, Theorem 7.1.8) $\phi=\sum_j \langle \phi,f^j\rangle f^j$ for any $\phi\in{\cal H}$. Orthonormal bases are automatically frames, but the
 converse is false; the frame property means simply that the sum of projectors $\sum_j \langle \cdot,f^j\rangle f^j$ is the identity operator, which is possible even when the $f^j$ are
 not linearly independent (typically, $1,e^{2\II\pi/3},e^{4\II\pi/3}\in\R^2$ make up a frame of $\R^2$). 

One checks immediately that the Fourier transformed family  $(\hat{f}^j)_{j\in I}$ is also a frame if ${\cal H}=L^2(\R^d)$.

\medskip
Higher-dimensional wavelets (say, living on $\R^d$, $d\ge 2$) depend on the choice of a scaling matrix $A$; if they are compactly supported on Fourier, their support must be an enlargement of
one of the fundamental domains of $\R^d$ under the action of $A$; in practice, say,  $|\xi|\in[2^{j-1},2^{j+1}]$ if $A=2\Id$. In this case there is only {\em one momentum scale}.
 On the other hand, in our case, we are interested in generalizations of wavelets having {\em several different} momentum scales. In the case of the construction by Cand\'es-Donoho-Demanet,
  the norm scale $j$ corresponds to
 $|\xi|$, and the angular  scale $j/2$ to the direction or angular sector cut along the sphere $|\xi|=2^j$. A direct tensorization of the one-dimensional construction is not possible since
 the angular scale depends on the norm scale. The authors obtain instead frames which are almost as good as actual wavelets, in the sense that the scalar product of two elements of the frame
 have a fast decrease  in the 'distance' (to be defined) between their windows. 

\medskip
As is often the case, we shall disregard the infra-red region $j\le 0$ and 'resum' the $(\psi^j)_{j\le 0}$ into a 'father wavelet' $\psi_0$ defined through its Fourier transform,
\BEQ \psi_{0}(\xi)=\sqrt{1-\sum_{j\ge 1} |\hat{\psi}(2^{-j}\xi)|^2}. \label{eq:father} \EEQ

%%%%%%%%%%%%%%%%%%

\subsection{The Cand\`es-Donoho-Demanet curvelet construction}

%%%%%%%%%%%%%%%%%%%%%%%%%

The following lines are not a  direct repetition of
the contents of \cite{CanDem,CanDon} since we are interested here in {\em three} space dimensions, but obviously the passage from 2d to 3d is very straightforward. The construction
relies mainly on two fundamental remarks:

\begin{Lemma} \label{lem:frame-construction}
\begin{itemize}
\item[(i)] (scaled Fourier window property) Assume that $(\hat{\nu}^{j,l})_{j,l}$ $(j\ge 1,-2^{-j/2}\le l<2^{j/2})$ is a set of $L^2$-functions such that for all $\xi\in\R^3$, $\sum_l |\hat{\nu}^{j,l}(\xi)|^2=1$, i.e.
satisfying a {\em scaled Fourier window property}. Then
$\sum_{j,l} |\hat{\psi}^j(\xi) \hat{\nu}^{j,l}(\xi)|^2=\sum_j |\hat{\psi}^j(\xi)|^2$. In particular, if $\psi$ is an orthonormal wavelet, hence satisfies the first basic wavelet equation
(\ref{eq:basic-eq}), then $\left( (\hat{\psi}^j\hat{\nu}^{j,l})_{j,l},\hat{\psi}_0\right)$ also satisfy the first basic wavelet equation. 
\item[(ii)] (space localization trick) Assume  $(f^j)_{j\in\N}$ is a family of functions in $L^2(\R)$ such that $\supp\,  (\hat{f}^j)\subset[0,2\pi\cdot 2^j]$ and, for all $\xi$,
 $\sum_j |\hat{f}^j(\xi)|^2=1$, i.e. satisfying the first basic wavelet equation. Then $f^{j,k}(x):=2^{-j/2} f^j(x-k2^{-j})$, $j\in\N,k\in\Z$ make up a frame of $L^2(\R)$.
 \end{itemize}
\end{Lemma}

{\em Proof.} (i) is trivial. One way to construct such functions $\hat{\nu}^{j,l}$ is to start from a smooth function $\hat{\nu}:\R\to\R$ with support in $[-2\pi,2\pi]$   such that
$|\hat{\nu}(\xi)|^2+|\hat{\nu}(\xi-2\pi)|^2=1$ for $0<\xi<2\pi$. Then an easy computation shows that $\sum_{-2^{j/2}\le l<2^{j/2}} |\hat{\nu}^{j,l}(\xi)|^2=1$ on $[0,2\pi]$, where
$\hat{\nu}^{j,l}(\xi):=\hat{\nu}(2^{j/2}(\xi-2\pi\cdot 2^{-j/2}l))$ is supported on $2\pi l 2^{-j/2}+[-2\pi 2^{-j/2},2\pi 2^{-j/2}]$. 

For (ii), note that $\langle \hat{\phi},\hat{f}^{j,k}\rangle_{L^2(\R)}=\langle \hat{\phi}\hat{f}^j,2^{-j/2}e^{-\II k2^{-j}\xi}\rangle_{L^2([0,2\pi\cdot 2^j])}$
by the support condition; thus, $\langle \hat{\phi},\hat{f}^{j,k}\rangle_{L^2(\R)}$ is the $k$-th Fourier coefficient of $\hat{\phi}\hat{f}^j$ viewed as a $2\pi\cdot 2^j$-periodic function.
Using the usual Parseval theorem for Fourier series and then the first basic wavelet equation, one obtains
\BEQ \sum_{j,k} |\langle \hat{\phi},\hat{f}^{j,k}\rangle|^2=\sum_{j} \int |\hat{\phi}\hat{f}^j(\xi)|^2 d\xi=\int |\hat{\phi}(\xi)|^2 d\xi.\EEQ
Hence $(\hat{f}^{j,k})_{j,k}$ make up a frame of $L^2(\R)$.      

\bigskip

Thus one easily cooks up a recipe for constructing a frame of "curvelets" $(\psi_{\Del})$ on $\R^3$ as follows. Recall from the preceding paragraph that there are two momentum scales, the norm scale $j$, and the
angular scale $j/2$.  The windows $\Del$ are now rotated rectangles of the form $\vec{x}_{\Del}+{\cal R}^{\vec{e}_{\Del}}([-2^{-j},2^{-j}]\times[-2^{-j/2},2^{-j/2}]^2)$, where
${\cal R}^{\vec{e}_{\Del}}$ is some rotation sending the vector $(1,0,0)$ into the unit vector $\vec{e}_{\Del}$ called {\em direction of the curvelet}. 
One requires that (i) the $\vec{e}_{\Del}$, $\Del\in\D^j$ are regularly scattered  $O(2^j)$ vectors on the unit sphere ${\cal S}^2:=\{\xi\in\R^3\ |\ |\xi|=1\}$, namely, $|\vec{e}_{\Del}-\vec{e}_{\Del'}|\approx 2^{-j/2}$ if
$\Del'\not=\Del$;  (ii) the Fourier support of the curvelets is 'dual' to their support in
direct space, i.e. $({\cal R}^{\vec{e}_{\Del}})^{-1}(\supp\,  \hat{\psi}_{\Del})\subset [2^{j-1},2^{j+1}]\times [- 2^{j/2}, 2^{j/2}]^2$.  
Here is a possible construction of a frame $(\psi_{\Del})$ satisfying the required properties:

\begin{itemize}
\item[(i)] (radial dependence)
Start from a Lemari\'e-Meyer wavelet on $\R$, $\psi$.
\item[(ii)] (angular dependence) Introduce two local maps on the sphere ${\cal S}^2$, 
Map:$\ U\stackrel{\sim}{\to} (-\pi,\pi)^2$ and $\overline{{\mathrm{Map}}}:\ \bar{U}\stackrel{\sim}{\to} (-\pi,\pi)^2$ (left composition of Map by the symmetry $(x_1,x_2,x_3)\mapsto (x_1,x_2,-x_3)$) defined on a neighbourhood $U$, resp. $\bar{U}=\{(x_1,x_2,-x_3)\ |\ (x_1,x_2,x_3)\in U\}$ of the North (resp. South)
hemisphere in ${\cal S}^2$, and two functions $I,\bar{I}:{\cal S}^2\to\R_+$ such that $\supp\,  I\subset U,\supp\,  \bar{I}\subset \bar{U}$ and $I^2+\bar{I}^2\equiv 1$.  These maps will allow us
to define curvelets  with direction centered around unit vectors $\vec{e}=\vec{e}_{l_1,l_2}={\mathrm{Map}}^{-1}(\pi l_1 2^{-j/2},\pi l_2 2^{-j/2})$, $-2^{j/2}\le l_1,l_2<2^{j/2}$ or the symmetric ones $\overline{\vec{e}}_{l_1,l_2}$. 
Let then $\hat{\nu}^{j,\vec{e}_{l_1,l_2}}(\theta_1,\theta_2)=\hat{\nu}^{j,l_1}(\theta_1)\hat{\nu}^{j,l_2}(\theta_2)$ be defined as in the proof of (i) in Lemma \ref{lem:frame-construction}, and
\BEQ \hat{\psi}^{j,\vec{e}}(\vec{\xi})=\hat{\psi}(2^j|\vec{\xi}|) I(\frac{\vec{\xi}}{|\vec{\xi}|}) \hat{\nu}^{j,\vec{e}}({\mathrm{Map}}(\frac{\vec{\xi}}
{|\vec{\xi}|})),  \qquad {\mathrm{resp.}}\qquad
\hat{\psi}^{j,\overline{\vec{e}}}(\vec{\xi})=\hat{\psi}(2^j|\vec{\xi}|) \bar{I}(\frac{\vec{\xi}}{|\vec{\xi}|}) \hat{\nu}^{j,\vec{e}}(\overline{\mathrm{Map}}(\frac{\vec{\xi}}
{|\vec{\xi}|}))   \label{eq:hat-psi-j-e} \EEQ
By Lemma \ref{lem:frame-construction} (i), the functions $(\psi^{j,\vec{e}},\psi^{j,\overline{\vec{e}}})_{j,\vec{e}}$ satisfy the first basic wavelet equation. In the sequel,
$\vec{e},$ Map, $U$ shall stand either for $\vec{e}=\vec{e}_{l_1,l_2}$, Map, $U$ or their symmetric counterparts $\overline{\vec{e}}$, $\overline{{\mathrm{Map}}}$, $\bar{U}$.

\item[(iii)] (space localization) Fix $j$ and $\vec{e}$. Then $\hat{\psi}^{j,\vec{e}}\circ {\cal R}^{\vec{e}}$ is supported in the anisotropic rectangle $[0,2^{j+1}]\times
[-C 2^{j/2},C 2^{j/2}]$ for some large enough constant $C$. By a simple rescaling one may assume that $C=1$.
Applying the space localization trick (see Lemma \ref{lem:frame-construction} (ii)) to the three space coordinates yields a frame $(\psi^{j,\vec{e},k})_{j,\vec{e},\vec{k}}$, $\vec{k}\in\Z^3$,
with \footnote{Note the scaling prefactor $2^{-j}=\sqrt{2^{-j}\times (2^{-j/2})^2}$. In a two-dimensional setting \cite{CanDem} one would find instead $2^{-3j/4}=\sqrt{2^{-j}\times
2^{-j/2}}$.}
\BEQ \psi^{j,\vec{e},\vec{k}}(\vec{x})=2^{-j} \psi^{j,\vec{e}}(\vec{x}-{\cal R}^{\vec{e}}(k_1 2^{-j},k_2 2^{-j/2},k_3 2^{-j/2})).\EEQ 

\end{itemize}

By construction the Fourier supports of the curvelets essentially do not overlap: $\langle \psi^{j,\vec{e},k},\psi^{j',\vec{e}',k'}\rangle=0$ except
if $j$ and $j'$, $\vec{e}$ and $\vec{e}'$ are {\em Fourier neighbours}, i.e. $|j-j'|\le 1$ and $|\vec{e}-\vec{e}'|\lesssim 2^{-j/2}$, a finite, $j$-independent number of possibilities for
$(j,\vec{e})$ fixed. On the other hand, the amplitude  $|\psi^{j,\vec{e},k}(x)|^2$ is 'maximal' on $\Del=\Del(j,\vec{e},k)={\cal R}^{\vec{e}}(
(k_1 2^{-j},k_2 2^{-j/2},k_3 2^{-j/2})+[-2^{-j},2^{-j}]\times[-2^{-j/2},2^{-j/2}]^2)$, a thin, flat piece of wall (for $j$ large) perpendicular to the curvelet direction 
$\vec{e}_{\Del}$.

Again  by construction, $|\hat{\psi}_{\Del}(\vec{\xi})|\lesssim 2^{-j_{\Del}}\approx \left(\Vol(\supp\,  \hat{\psi}_{\Del}) \right)^{-\half}$ and (as a simple variable rescaling shows)
$|\psi_{\Del}(x)|\lesssim 2^{j_{\Del}}\approx (\Vol(\Del))^{-\half}$. Note that $\Vol(\supp\,  \hat{\psi}_{\Del})\Vol(\Del)\approx 1$, an optimal upper bound in view of the uncertainty 
principle.

%%%%%%%%%%%%%%%%%%%%%%%%%%

\subsection{Minkowski curvelets}

%%%%%%%%%%%%%%%%%%%%%%%%%%%%%%%%%

The previous curvelets are a nice tool for the study of the solution of a wave equation at {\em fixed time} $t$. On the other hand, controling e.g. $||G(F)||_{L^p_t(L^q_{\vec{x}})}$
in view of Strichartz estimates requires the introduction of a new type of {\em space-time curvelets} $(\psi_{\Del})$ essentially supported inside a compact {\em space-time window},
and such that $G(t,\vec{x};t',\vec{x}')$ is very small when $x$ and $y$ belong to far away windows. This is a priori contradictory with the fact that waves propagate. In order
that $G(t,\vec{x};t',\vec{x}')$ be small for $\vec{x}'\approx \vec{x}(t)$ following the semi-classical Hamiltonian flow, one must impose a Fourier support condition on
$\psi_{\Del}$ which contradicts propagation.

Let us proceed a little informally at this stage. Replace $\Box$ by the usual wave operator $\partial_t^2-\Del$, so that $G$ is the convolution kernel
\BEQ G(t,\vec{x})=\int \frac{\sin t|\vec{\xi}|}{|\vec{\xi}|} e^{\II \vec{x}\cdot \vec{\xi}}d\vec{\xi}, \EEQ
see e.g. \cite{Sog}, or (in relativistic coordinates)
\BEQ G(t,\vec{x})=\frac{\II}{2} \left\{ \int \del(\xi_0-|\vec{\xi}|)\frac{e^{-\II x\cdot \xi}}{|\vec{\xi}|} d\xi -  \int \del(\xi_0+|\vec{\xi}|)\frac{e^{-\II x\cdot \xi}}{|\vec{\xi}|} d\xi \right\}.
\label{eq:del} \EEQ
The latter expression emphasizes the fact that the essential contribution to\\ $\hat{G}(\xi)=\lim_{\eps\to 0^+} \frac{1}{\xi_0^2-|\vec{\xi}|^2-\II \sgn(\xi_0)\eps}$ comes from the region $\xi_0=\pm|\vec{\xi}|$.
Assume for the sake of the discussion that $\xi_0>0$. In order to contradict propagation, it is natural to require that $\xi_0-|\vec{\xi}|$ be bounded away from $0$. Thus one is led
to introduce two scales $m\in\Z$ and $j\in\N$ and a corresponding smooth cut-off $\hat{\chi}^{m,j}(\xi)$ for which 
\BEQ |\xi_0-|\vec{\xi}||\approx 2^{m-j/2}, \quad |\vec{\xi}|\approx 2^{m+j/2}.\EEQ
In the following discussion we assume $j\ge 1$ so that $|\xi_0|\approx |\vec{\xi}|$.  These constraints are preserved by transverse variations
 $\del \vec{\xi}_{\perp}\perp\vec{\xi}$ of $\vec{\xi}$ of order $|\del\vec{\xi}_{\perp}|\lesssim 2^m$ since
 \BEQ |\vec{\xi}+\del\vec{\xi}_{\perp}|-|\vec{\xi}|=|\vec{\xi}|\left( \sqrt{1+|\del\vec{\xi}_{\perp}|^2/|\vec{\xi}|^2}\right)\approx \half \frac{|\del\vec{\xi}_{\perp}|^2}{|\vec{\xi}|}
 \lesssim 2^{-m-j/2} \label{eq:sqrt} .\EEQ

We now consider the bulk of the support of $G$ in space-time. Assume to simplify that $\vec{x}$ is parallel to $\vec{\xi}$. Then $e^{-\II x\cdot \xi}$ in eq. (\ref{eq:del})
may be rewritten as $e^{-\II \left[ (\xi_0-|\vec{\xi}|)x^0+|\vec{\xi}|(x^0-|\vec{x}|)\right]}$. Oscillations due to the phase imply fast decay of $\chi^{m,j}\ast G \ast \chi^{m,j}$
when $|x^0|\ll (\xi_0-|\vec{\xi}|)^{-1}\approx 2^{-m+j/2}$ or $|x^0-|\vec{x}||\ll |\vec{\xi}|^{-1}\approx 2^{-m-j/2}$. In particular, this means that waves propagate only up to a
propagation time of order $O(2^{-m+j/2})$.

 \bigskip
 The construction of  Minkowski curvelets obeys the following principles. Since $|\del\vec{\xi}_{\perp}|/|\vec{\xi}|\approx 2^{-j/2}$ at most, it is clear that a proper Fourier space decomposition should include
 an angular decomposition of $\vec{\xi}$ with $O(2^j)$ sectors, in precise analogy with the case of curvelets, up to a rescaling by the {\em lower Fourier scale} $|\xi_0-|\vec{\xi}||\approx
 2^{m-j/2}$, namely, $\frac{|\vec{\xi}|}{|\xi_0-|\vec{\xi}||}\approx 2^j$ and $\frac{|\del\vec{\xi}_{\perp}|}{|\xi_0-|\vec{\xi}||}\approx 2^{j/2}$.
Correspondingly, the shape of a {\em Minkowski curvelet} in {\em real space} should be similar to that of {\em curvelets} up to rescaling by $|x^0|$, i.e. the space-time coordinate
dual to $\xi_0-|\vec{\xi}|$. 

\medskip

Let us now be a little more precise. The inverse Fourier transform
yields the exponential $e^{-\II x\cdot \xi}$, with $x\cdot \xi=x_0\xi_0-\vec{x}\cdot \vec{\xi}$. Assume the direction of propagation $\vec{e}_{\Del}$ is along the first axis of coordinates,
$\vec{e}_{\Del}=(1,0,0)$. Then $x\cdot\xi\approx x_0\xi_0-x_1\xi_1=x_0(\xi_0-\xi_1)-(x_1-x_0)\xi_1\approx x_0(\xi_0-|\vec{\xi}|)-(|\vec{x}|-x_0)|\vec{\xi}|$. Since $\xi_0-|\vec{\xi}|\approx  2^{m-j/2}$
and $\xi_0\approx |\xi|\approx  2^{m+j/2}$, one obtains by dualizing $|x_0|, |\vec{x}|\lesssim 2^{j/2-m}$  and 
$|x_0-|\vec{x}||\lesssim 2^{-j/2-m}$.  Assume $|\vec{x}|\approx 2^{j/2-m}$, which is generically the case.   The constraint $|x_0-|\vec{x}||\lesssim 2^{-j/2-m}\ll
|\vec{x}|\approx 2^{j/2-m}$ imposes (as in eq. (\ref{eq:sqrt}) above) $|x_i|\lesssim 2^{-m}$, $i=2,3$ for transverse coordinates, dual to the condition $|\xi_i|\lesssim  2^m$.
In other words, rescaling by the typical time of propagation $|x_0|\approx ^{-1} 2^{j/2-m}$ and replacing $x_1$ by the proper invariant coordinate along
light rays, $x_1-x_0$, i.e. letting $y_1:=2^{m-j/2}(x_1-x_0)$ and $y_i=2^{m-j/2} x_i$, $i=2,3$ for transversal coordinates, one has obtained  $|y_1|\lesssim 2^{-j}$ and
 $y_i\lesssim 2^{-j/2}$,  $i=2,3$ as for Cand\`es-Donoho-Demanet curvelets.

Summarizing, the window of a Minkowski curvelets $\psi_{\Del}$ should be a rotated 4d rectangle oriented along the four-vector
$e_{\Del}=(1,\vec{e}_{\Del})$, 
\BEA &&  \Del={\cal R}^{\vec{e}}\left( \left\{ z_0 2^{j/2-m} (1,1,0,0)+2^{-m} (0,0,z_2,z_3)+ z_1 2^{-j/2-m} (1,-1,0,0);\right.\right.\nonumber\\
&& \left.\left. \qquad \qquad \qquad \qquad \qquad  k_i-1\le z_i\le k_i+1,\ i=0,\ldots,3\right\}
\right),\EEA
where the 4-vector $(1,1,0,0)$ is in the direction of propagation, $(0,0,z_2,z_3)$ is orthogonal to the propagation plane span$((1,0,0,0),e_{\Del})$,  whereas $(1,-1,0,0)$ is some arbitrary transverse direction, here chosen to be the symmetric null direction in
the propagation plane.

\begin{figure}[h]
  \centering
   \includegraphics[scale=0.70]{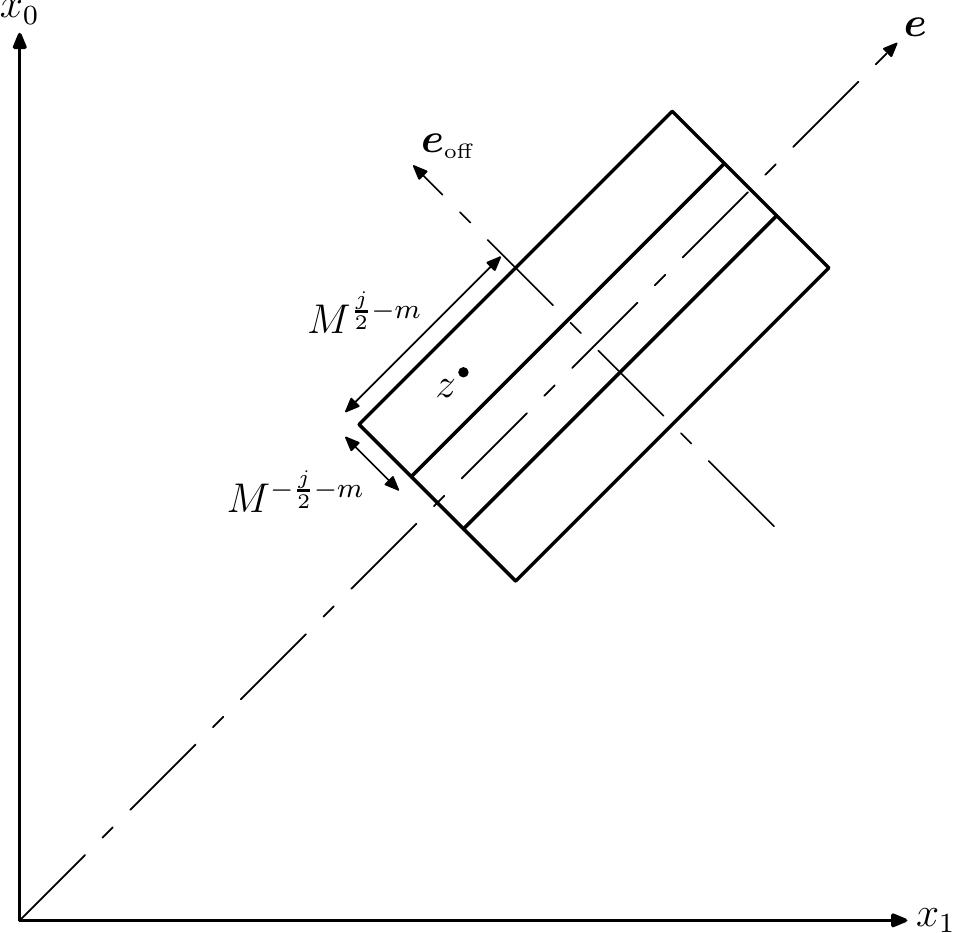}
   \caption{\small{$(M=2)$. Space-time tiling for a propagation along the first axis. Transversal directions, $e_{\perp}=(0,0,1,0)$ or $(0,0,0,1)$ are perpendicular
   to the plane of propagation. The size of the window along these two directions is $2^{-m}$. }}
  \label{Fig1}
\end{figure}

\begin{figure}[h]
  \centering
   \includegraphics[scale=0.70]{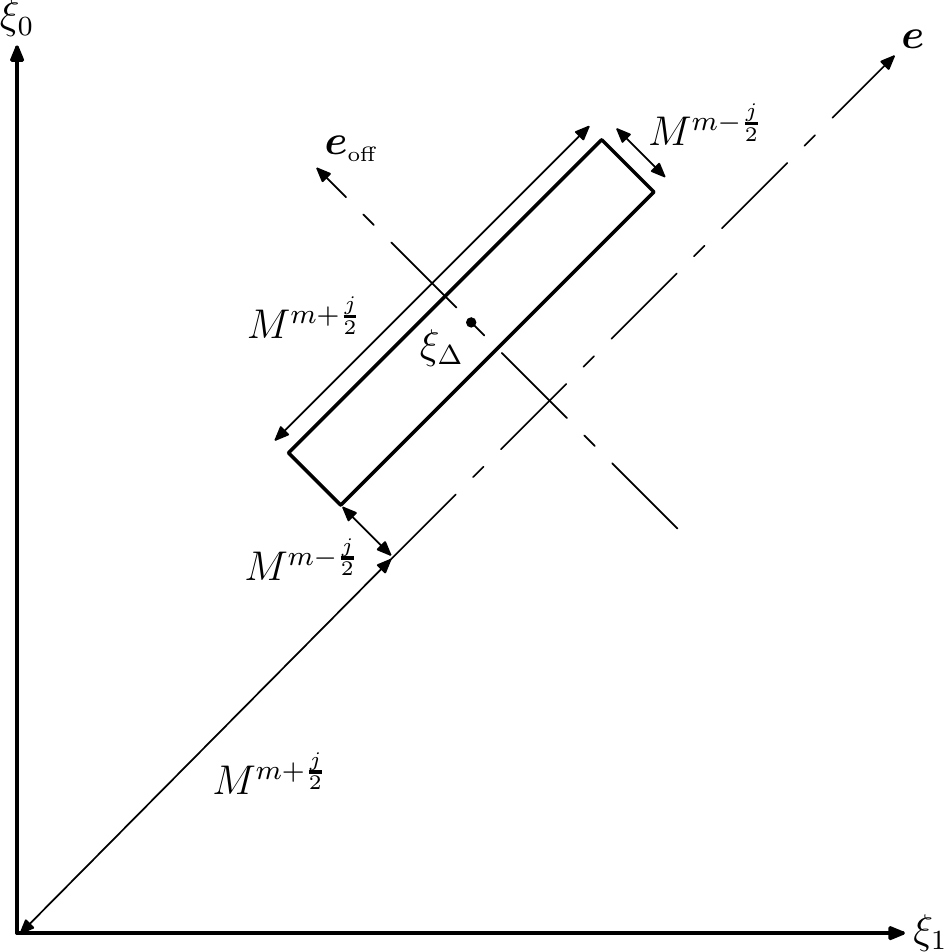}
   \caption{\small{$(M=2)$. Support of a curvelet in Fourier space. Respective scalings are $e_{\Del}\cdot \xi\approx 2^{m-j/2}$, $\xi_0\approx |\vec{\xi}|\approx 2^{m+j/2}$
   and $0\le |\vec{\xi}|\lesssim 2^m$ for transversal directions.}}
  \label{Fig2}
\end{figure}

The following easy adaptation of the construction of the previous paragraph defines a {\em frame of  Minkowski curvelets} satisfying the previous requirements:

\begin{itemize}
\item[(i)] (Fourier scales) start from a one-dimensional Lemari\'e-Meyer wavelet, $\psi$,  and let 
\BEQ \hat{\psi}^{m,j}(\xi):=\hat{\psi}(\frac{\sqrt{|\xi^2|}}{ 2^m})
\hat{\psi}(\frac{\xi_0^2/|\xi^2|}{2^j})  \qquad (j\ge 1),\label{eq:LMM} \EEQ
\BEQ \hat{\psi}^m_0(\xi)=\hat{\psi}(\frac{\sqrt{|\xi^2|}}{ 2^m}) \hat{\psi}_0(\xi_0^2/|\xi^2|) \EEQ
where $\hat{\psi}_0$ is the 'father wavelet' defined in eq. (\ref{eq:father}).

 If $\xi\in\supp\, (\hat{\psi}^{m,j})$ and, say, $\xi_0>0$, then 
 \BEQ |\xi^2|\approx  2^{2m}, \qquad ||\xi_0|-|\vec{\xi}||\approx 2^{m-j/2}, \qquad |\xi_0|\approx |\vec{\xi}|\approx 2^{m+j/2} \label{eq:1.13} \EEQ as required above.
On the other hand, the support of $\hat{\psi}^m_0$ is made up of 'elliptic' regions which are irrelevant for geometrical optics where either $|\xi_0|\ll |\vec{\xi}|$ or $|\vec{\xi}|\ll |\xi_0|$. 
Estimates are very easy in these and we shall not bother any longer about $\hat{\psi}_0^m$ in the remainder of the article.

\item[(ii)] (angular dependence) Let (replacing simply $\hat{\psi}^j$ by $\hat{\psi}^{m,j}$ in eq. (\ref{eq:hat-psi-j-e}))
\BEQ \hat{\psi}^{m,j,\vec{e}}(\xi)=\hat{\psi}^{m,j}(\xi) I(\frac{\vec{\xi}}{|\vec{\xi}|}) \hat{\nu}^{j,\vec{e}}({\mathrm{Map}}(\frac{\vec{\xi}}
{|\vec{\xi}|})).  \label{eq:1.12} \EEQ
Hence (decomposing $\xi\in \supp\, (\hat{\psi}^{m,j,\vec{e}})$ into $\left(\xi_0,(\vec{e}\cdot \vec{\xi})\vec{\xi}+\vec{\xi}_{\perp}\right)$) there are {\em three Fourier scales}: the {\em lower Fourier scale}, $|\xi_0-|\vec{\xi}||\approx |\xi_0-\vec{e}\cdot \vec{\xi}|\approx  2^{m-j/2}$; the
{\em middle one}, $|\vec{\xi}_{\perp}|\approx  2^m$; and the {\em higher one}, $\xi_0\approx |\xi|\approx \xi_0+|\xi|\approx  2^{m+j/2}$. If $\vec{e}=(1,0,0)$, then
$\supp\, (\hat{\psi}^{m,j,\vec{e}})$ is included in some enlargement of the rectangle defined by $|\xi_1-\xi_0|\le  2^{m-j/2}$, $|\xi_2|,|\xi_3|\le  2^m$, $|\xi_0|\le  2^{m+j/2}$.
As for the real-space window $\Del$, the Fourier support is oriented along the four-vector $e_{\Del}$. It is shifted along that same direction, so that 
$0\not \in \supp\, (\hat{\psi}^{m,j,\vec{e}})$.

\item[(iii)] The light cone and the removal of the irrelevant zone $|\xi_0|\ll |\vec{\xi}|$ cut the remaining Fourier domain into four connected components, depending on the
signs of $\xi_0$ and $|\xi_0|-|\vec{\xi}|$. Thus $\hat{\psi}^{m,j,\vec{e}}$ may be rewritten as the sum of four functions with disjoint supports, which we still denote by 
$\hat{\psi}^{m,j,\vec{e}}$ to keep notations reasonable. 

\item[(iv)] (space localization) Let
\BEA &&  \psi^{m,j,\vec{e},\vec{k}}(\vec{x})=2^{-2m} \psi^{m,j,\vec{e}}\left(x-
      {\cal R}^{\vec{e}}\left( k_0 2^{j/2-m} (1,1,0,0)+ \right. \right. \nonumber\\ && \left.\left. \qquad \qquad \qquad \qquad 
      2^{-m} (0,0,k_2,k_3)+k_1 2^{-j/2-m} (1,-1,0,0) \right)\right). \label{eq:1.16} \EEA 
\end{itemize}

We let $\D$ be the set of all Minkowski curvelet indices $(m,j,\vec{e},\vec{k})$. If $m$ is fixed, then the set $\D$ restricts to $\D^m$; if $m,j$ are fixed, then $\D$ restricts to
$\D^{m,j}$, and similarly for $\D^{m,j\vec{e}}$. The {\em Fourier support} of $\Del$ is $\supp\, _{\xi}(\Del):=\supp\, (\hat{\psi}_{\Del})$ (it does not depend on $\vec{k}$). We also
choose $x_{\Del}\in\Del$ for each $\Del$, for instance 
\BEQ x_{\Del}={\cal R}^{\vec{e}}\left( k_0 2^{j/2-m} (1,1,0,0)+ 
      2^{-m} (0,0,k_2,k_3)+k_1 2^{-j/2-m} (1,-1,0,0)\right). \EEQ

Note that for Minkowski curvelets,
\BEQ \Vol(\supp\, _{\xi}(\Del))\approx 2^{4m}\approx (\Vol(\Del))^{-1}.\EEQ

%%%%%%%%%%%%%%%%%%%%%%%%%%%%%%
%%%%%%%%%%%%%%%%%%%%%%%%%%%%

\section{Decay estimates for Minkowski curvelets}

%%%%%%%%%%%%%%%%%%%%%%%%%%%
%%%%%%%%%%%%%%%%%%%%%%%%%%

We shall now show estimates for $\psi_{\Del}, \hat{\psi}_{\Del}$ and for  the scalar products $\langle \psi_{\Del},\psi_{\Del'}\rangle$, showing in particular a summable, fast  decay
in the scaled distance between $\Del$ and $\Del'$, in the spirit of \cite{CanDem}.

%%%%%%%%%%%%%%%%%%%%%%%%%%%%%%%

%%%%%%%%%%%%%%%%%%%%%%%%%%%%%%

As in the Cand\`es-Donoho-Demanet construction, we now define {\em Minkowski molecules}, which have the same real-space and Fourier space decay as Minkowski curvelets. 
We need a good choice of local coordinates for a fixed unitary vector $\vec{e}=\vec{e}_{\Del}$.  Let first  $(\vec{e},\vec{e}_{\perp,1},\vec{e}_{\perp,2})={\cal R}^{\vec{e}}
((1,0,0),(0,1,0),(0,0,1))$ be an orthonormal basis of $\R^3$, and $(x_{//},\vec{x}_{\perp}=(x_{\perp,1},x_{\perp,2}))$ the dual coordinates. Then one must choose an adapted basis of vector fields;
we let $\nabla_{//}=\nabla_e=\partial_{x^0}+\partial_{x_{//}}$; $\nabla_{\perp,i}=\partial_{x_{\perp,i}}$, $i=1,2$ be a basis of the transversal plane ($\nabla_{\perp}$
stands for some linear combination of $\nabla_{\perp,i}$, $i=1,2$); and $\nabla_{off}$  be any derivative in directions away from the propagation plane, for instance
$\nabla_{off}=\partial_{x_0}-\partial_{x_{//}}$. Note that $e\cdot x=x^0-x_{//}$ (Minkowski scalar product) gives apparently the 'component' of $x$ along the direction
$(1,-\vec{e}_{\Del})$ which is {\em 'perpendicular'} to the direction of propagation; but both words 'component' and 'perpendicular' refer to the Euclidean scalar product on $\R^4$, which
is totally misleading in this context. On the other hand, $e\cdot x$ is not a (Minkowski) 'projection' of $x$ along $e$ either since $e\cdot e=0$. In any case, $e\cdot x$ should not be
mistaken with $\vec{e}\cdot \vec{x}$, which is the Euclidean projection of $\vec{x}\in\R^3$ in the direction $\vec{e}$. We also let $e_{\Del}^*=e_{\Del}$ be Minkowski orthogonal to $e_{\Del}$,
and $\nabla_{//}^*=\nabla_{e_{\Delta}^*}=\nabla_{//}$
(in the next paragraph, $e_{\Del}$ will not be a null vector any more, and we will have $e_{\Del}^*\not=e_{\Del}$, $\nabla_{//}^*\not=\nabla_{//}$).

\begin{Definition}[Minkowski molecules] \label{def:molecules}
A family $(a_{\Del})_{\Del\in\D}$ of $C^N$ functions is a {\em family of Minkowski molecules} of regularity $N$ if
\BEQ a_{\Del}(x)=f_{\Del}\left(x-{\cal R}^{\vec{e}_{\Del}}\left(k_0 2^{j/2-m}(1,1,0,0)+2^{-m}(0,0,k_2,k_3)+k_1 2^{-j/2-m} (1,-1,0,0)\right)\right),\EEQ
where $f_{\Del}$ satisfies the following real-space and Fourier space decays,
\begin{itemize}
\item[(i)] If $r=r_{//}+r_{\perp}+r_{off}\le N$,
\BEA && |(\nabla_{//}^*)^{r_{//}}\nabla_{\perp}^{r_{\perp}}\nabla_{off}^{r_{off}}f(x)|\lesssim 2^{2m}\cdot \nonumber\\
 && \qquad \qquad \qquad (2^{m-j/2})^{r_{//}} (2^m)^{r_{\perp}} (2^{m+j/2})^{r_{off}} \cdot \nonumber\\
 && \qquad \qquad \qquad \qquad \left( \langle 2^{m+j/2} e_{\Del}\cdot x\rangle + \langle 2^m |\vec{x}_{\perp}| \rangle +\langle 2^{m-j/2} |x^0|\rangle \right)^{-N}. \label{eq:space-decay}
 \EEA
The decay is strongest 'along' $e_{\Del}$, i.e. for $e_{\Del}\cdot x$ (away from the direction of propagation), weaker in the two transverse directions, and still weaker in arbitrarily
chosen directions; in (\ref{eq:space-decay}), $|\vec{x}_{\perp}|$, resp. $|x^0|$ may be replaced by any linear combination $|\vec{x}_{\perp}|+C|e_{\Del}\cdot x|$, resp. $|x^0|+C'|\vec{x}_{\perp}|+C'' |e_{\Del}
\cdot x|$.
\item[(ii)] (Fourier decay)
\BEQ |\hat{f}_{\Del}(\xi)|\lesssim 2^{-2m} \left( \langle\langle 2^{-m+j/2} e_{\Del}^*\cdot \xi\rangle\rangle + \langle 2^{-m} |\vec{\xi}_{\perp}|\rangle+
\langle\langle 2^{-m-j/2} \xi_0 \rangle\rangle \right)^{-N}  \label{eq:Fourier-decay} \EEQ
(recall $\langle\langle t\rangle\rangle=1+t+t^{-1}$). Here again, the decay is strongest 'along' $e_{\Del}$, weaker in the two transverse directions, and still weaker in arbitrarily chosen
directions; in (\ref{eq:Fourier-decay}), $|\vec{\xi}_{\perp}|$, resp. $|\vec{\xi}^0|$ may be replaced by any linear combination $|\vec{\xi}_{\perp}|+C|e_{\Del}\cdot \xi|$,
 resp. $|\xi^0|+C'|\vec{\xi}_{\perp}|+C'' |e_{\Del}\cdot \xi|$. Note that $|e\cdot \xi|$ and $|\xi_0|$ are bounded from below, while $|\vec{\xi}_{\perp}|$ is not.
\end{itemize}
\end{Definition}

\begin{Lemma} \label{lem:mol}
Minkowski curvelets are Minkowski molecules. Furthermore, the following additional Fourier decay holds for centered curvelets. 
Assume $\Del=(m,j,\vec{e},(0,0,0,0))$ is centered at the origin of space-time. Then (\ref{eq:Fourier-decay}) extends to derivatives of $\hat{\psi}_{\Del}$,
\BEA && |\nabla_{//}^{r_{//}}\nabla_{\perp}^{r_{\perp}}\nabla_{off}^{r_{off}}\hat{\psi}_{\Del}(\xi)|\lesssim 2^{-2m} \cdot \nonumber\\
 && \qquad \qquad \qquad (2^{-m-j/2})^{r_{//}} (2^{-m})^{r_{\perp}} (2^{-m+j/2})^{r_{off}} \cdot \nonumber\\
 && \qquad \qquad \qquad \qquad \qquad \left( \langle\langle 2^{-m+j/2} e_{\Del}^*\cdot \xi\rangle\rangle + \langle 2^{-m} |\vec{\xi}_{\perp}|\rangle+
\langle\langle 2^{-m-j/2} \xi_0 \rangle\rangle \right)^{-N}  \label{eq:Fourier-decay-centered}  \nonumber\\ \EEA

\end{Lemma}

{\bf Proof.} 
The Fourier decay is a trivial consequence of the prefactor $2^{-2m}$ in eq. (\ref{eq:1.13}) and the boundedness of Fourier support of $\psi_{\Del}$, see
eq. (\ref{eq:LMM}) and (\ref{eq:1.12}). Fourier transforming $\hat{\psi}_{\Del}$ and rescaling variables yields the inverse prefactor
$2^{-2m}\cdot \Vol_{\xi}(\Del)\approx 2^{2m}$. The differential operator $(\nabla_{//}^*)^{r_{//}}\nabla_{\perp}^{r_{\perp}}\nabla_{off}^{r_{off}}$ brings down a factor of order
$(\xi_0-\xi_{//})^{r_{//}} \xi_{\perp}^{r_{\perp}} \xi_0^{r_{off}}\approx (2^{m-j/2})^{r_{//}} (2^m)^{r_{\perp}} (2^{m+j/2})^{r_{off}} $ as in eq. (\ref{eq:space-decay}). Finally, if, say,
$e_{\Del}=(1,0,0)$, then
\BEA && \psi_{\Del}(x)=2^{2m}\int \hat{\psi}(\frac{\zeta^2}{^2}) \hat{\psi}(\zeta_0^2/\zeta^2) I(\vec{\zeta}) \hat{\nu}^{j,e_{\Del}}({{\mathrm{Map}}}(\vec{\zeta})) \nonumber\\ && \qquad
e^{\II [ 2^{m-j/2} (\zeta_0-\zeta_1)x^0-2^{m+j/2} \zeta_1 (x^1-x^0)-2^m(\zeta_2 x^2+\zeta_3 x^3)]} d\zeta \EEA
in rescaled coordinates $\zeta$, where all the functions $\hat{\psi}(\frac{\zeta^2}{^2})$, $\hat{\psi}(\zeta_0^2/\zeta^2)$,
 $ I(\vec{\zeta})$, $\hat{\nu}^{j,e_{\Del}}({{\mathrm{Map}}}(\vec{\zeta})$ have bounded, $j$- and $m$-independent support in $\zeta$, and are $C^{\infty}$ with scale-independent
 $||\ \cdot \ ||_{\infty}$-norms on their derivatives. The real-space decay in eq. (\ref{eq:space-decay}) is now obtained by differentiating a number of times
  with respect to $\zeta$ and integrating by parts.
  
  \medskip

Let us finally prove the additional Fourier decay for centered curvelets. Since $k_{\Del}=(0,0,0,0)$, $\psi_{\Del}=2^{-2m}\psi^{m,j,\vec{e}}$ with $m=m_{\Del},j=j_{\Del},\vec{e}=
\vec{e}_{\Del}$. The scaling behaviour of $\hat{\psi}^{m,j,\vec{e}}$ is the same as that of $\hat{\psi}^{m,j}(\xi)=\hat{\psi}(\frac{\sqrt{|\xi^2|}}{2^m})
\hat{\psi}(\frac{\xi_0^2/|\xi^2|}{2^j})$. Elementary computations yield
\BEQ |\nabla_{//}\hat{\psi}^{m,j}(\xi)|\lesssim \max\left( \frac{|\xi_0-\xi_{//}|}{2^{2m}}, \frac{\xi_0^2/\xi^2}{2^j}, \frac{\xi_0^2 |\xi_0-\xi_{//}|/\xi^4}{2^j} \right)
\lesssim 2^{-m-j/2} \EEQ and similarly
 \BEQ |\nabla_{off}\hat{\psi}^{m,j}(\xi)|\lesssim \max\left( \frac{|\xi_0+\xi_{//}|}{2^{2m}}, \frac{\xi_0^2/\xi^2}{2^j}, \frac{\xi_0^2 |\xi_0+\xi_{//}|/\xi^4}{2^j} \right)
\lesssim 2^{-m+j/2}; \EEQ
\BEQ |\nabla_{\perp} \hat{\psi}^{m,j}(\xi)|\lesssim \max\left(\frac{|\xi_{\perp}|}{2^{2m}},\frac{\xi_0^2 |\xi_{\perp}|/\xi^4}{2^j} \right)\lesssim 2^{-m}. \EEQ
More generally, each derivative $\nabla$, $\nabla=\nabla_{//},\nabla_{\perp},\nabla_{off}$ produces a supplementary scaling factor $2^{-m-j/2}$, resp. $2^{-m}, 2^{-m+j/2}$.

\hfill \eop

\bigskip

With this definition, Minkowski molecules (hence in particular Minkowski curvelets) are {\em almost orthogonal}, in the following sense.

\begin{Lemma}[almost-orthogonality of Minkowski molecules] \label{lem:almost-ortho}

Let $(a_{\Del})_{\Del\in\D}, (a'_{\Del})_{\Del\in\D}$ be two families of Minkowski molecules of regularity $4N$. Then
\BEQ |\langle a_{\Del},a'_{\Del'}\rangle|\lesssim d(\Del,\Del')^{-N},\EEQ
where 
\BEA  && d(\Del,\Del'):=2^{4|m-m'|+2|j-j'|} (1+|m-m'|^2)(1+|j-j'|^2) \nonumber\\ && \qquad \qquad \qquad  \left\{ 1+ d_{ang}(\Del,\Del')+d_{x,//}(\Del,\Del')+d_{x,\perp}(\Del,\Del')+d_{x,off}(\Del,\Del') \right\};  \label{eq:almost-ortho} \nonumber\\ \EEA

$d_{ang}(\Del,\Del'):=2^{\half \min(j,j')} |\vec{e}_{\Del}-\vec{e}_{\Del'}|$ is the rescaled angular distance between the Fourier supports $\supp\, _{\xi}(\Del)$ and $\supp\, _{\xi}(\Del')$, measured
in terms of the greater angular span of the two;

$d_{x,//}(\Del,\Del'):={\bf 1}_{m+j/2\le m'+j'/2} 2^{m+j/2} |e_{\Del}\cdot (x_{\Del}-x_{\Del'})| + {\bf 1}_{m'+j'/2\le m+j/2} 2^{m'+j'/2} |e_{\Del'}\cdot (x_{\Del'}-x_{\Del})|$ is a scaled
distance 'along' the direction of propagation;

$d_{x,\perp}(\Del,\Del'):={\bf 1}_{m\le m'} 2^m |\vec{e}_{\Del} \wedge (\vec{x}_{\Del}-\vec{x}_{\Del'})|+ {\bf  1}_{m'\le m} 2^{m'} |\vec{e}_{\Del'} \wedge (\vec{x}_{\Del'}-\vec{x}_{\Del})|$
is a scaled transversal distance;

$d_{x,off}(\Del,\Del'):=2^{\min(m-j/2,m'-j'/2)} |x_{\Del}^0 -x_{\Del'}^0|$ is a scaled distance in some direction away from the propagation plane.

\end{Lemma}

As in  Definition \ref{def:molecules}, $d_{x,\perp}$ and $d_{x,off}$ may also be redefined by using the  coordinates along somewhat arbitrary directions.

\medskip

{\bf Proof.} 
If $\Del,\Del'\in\D^{m,j}$ have same scale indices, then the decay in $d_{ang}$ follows from eq. (\ref{eq:Fourier-decay}) and the decay in $d_{x,//},d_{x,\perp},d_{x,off}$ from
eq. (\ref{eq:space-decay}). Note that $2^{2m}\approx 1/\sqrt{\Vol(\Del)}$ and $2^{-2m}\approx 1/\sqrt{\Vol(\supp\, _{\xi}(\Del))}$, hence $|\langle a_{\Del},a'_{\Del'}\rangle|$ is globally
of order $O(1)$.

In the general case, the various distances $d_{ang}, d_{x,//}, d_{x,\perp}, d_{x,off}$ should be measured in terms of the lower scale, $\min(k,k')$ ,where  $(k,k')=(j/2,j'/2),
(m+j/2,m'+j'/2)$, etc. because the window $\Del$ is not well localized in space (i.e. has rescaled size $\gg 1$) when measured in terms of the width $O(2^{-k'})$ of $\Del$ if
$k'>k$. Finally, factors $2^{-|m-m'|}$, resp. $2^{-|(m-j/2)-(m'-j'/2)|}$, $2^{-|(m+j/2)-(m'+j'/2)|}$ to arbitrary order $\le N$ may be derived from (\ref{eq:Fourier-decay}), yielding in particuliar
the required prefactor $\left( 2^{4|m-m'|+2|j-j'|} (1+|m-m'|^2)(1+|j-j'|^2) \right)^{-N}.$

 \hfill \eop

\begin{Lemma}[decay summability] \label{lem:decay-sum}
\begin{itemize}
\item[(i)]
There exists a constant $C$ such that, for every $r>5$ and $\Del,\Del''\in\D$, 
\BEQ \sum_{\Del'\in\D} d(\Del,\Del')^{-r} d(\Del',\Del'')^{-r}\le Cd(\Del,\Del'')^{-r}. \EEQ \label{eq:decay-summability}
\item[(ii)] (same hypotheses). There exists a constant $C>0$ such that, for every $\Del\in\D$,
\BEQ \sum_{\Del'\in\D} d(\Del,\Del')^{-r}\le C.\EEQ
\end{itemize}
\end{Lemma}

{\bf Proof.}

We shall only prove (i), since (ii) is simpler and may be proved along the same lines.
\begin{itemize}
\item[(i)] Assume first that $m=m''$, $j=j''$, $e_{\Del}=e_{\Del''}$, and sum over $\Del'\in \D^{m,j,e_{\Del}}$ only. Then 
\BEA &&  d(\Del,\Del')\approx 1+\max\left(2^{m+j/2}|e_{\Del}\cdot (x_{\Del}-x'_{\Del})|, 2^m |\vec{e}^{\Del}_{\perp,1} \cdot (\vec{x}_{\Del}-\vec{x}_{\Del'}) |, 
 \right.\nonumber\\ && \left. \qquad \qquad \qquad \qquad
 2^m |\vec{e}^{\Del}_{\perp,2} \cdot (\vec{x}_{\Del}-\vec{x}_{\Del'}) |,
2^{m-j/2} |x^0_{\Del}-x^0_{\Del'}| \right). \EEA

 Changing the real space localization of $\Del'$ essentially amounts to shifting one of these four
scaled quantities by an integer. Hence $\sum_{\Del'\in\D^{m,j,e_{\Del}}} d(\Del,\Del')^{-r} d(\Del',\Del'')^{-r}\le Cd(\Del,\Del'')^{-r}$ if and only if
the following holds,
\BEQ \sum_{x'} (C+||x-x'||)^{-r} (C+||x'-x''||^{-r}\lesssim (C+||x-x''||)^{-r}  \label{eq:Z4} \EEQ
if  $x,x'\in\Z^4$, $x'$ ranges over the integer lattice $\Z^4$ and $||\ \cdot \ ||$ is the Euclidean norm on $\R^4$. This inequality is true if $r>4$ for
\BEA && \int_{1<||x'-x||<\half ||x''-x||} (1+||x-x'||)^{-r} (1+||x'-x''||)^{-r} dx' \nonumber\\ && \qquad \qquad \approx (1+||x-x''||)^{-r} \inf_{1<||x'-x||<\infty} (1+||x-x'||)^{r} dx' \EEA
and the last integral is convergent.  

\item[(ii)] Assume now simply that $m=m''$ and $j=j''$, and sum over $\Del'\in\D^{m,j}$. Fix $\vec{e}_{\Del}$, say, $\vec{e}_{\Del}=(1,0,0)$. Index $\vec{e}_{\Del'}$ by
a couple of integers $\vec{l}=(l_1,l_2)$, $|l_1|,|l_2|\lesssim 2^{j/2}$, so that $|\vec{e}_{\Del}-\vec{e}_{\Del'}|\approx 2^{-j/2} |\vec{l}|$.  Then
\BEA &&  d_{x,//}(\Del,\Del')+d_{x,\perp}(\Del,\Del')+d_{x,off}(\Del,\Del')\approx  2^{m+j/2} |e_{\Del}\cdot (x_{\Del}-x_{\Del'})|+
\nonumber\\ && \qquad |\vec{l}| \left(
 2^m |\vec{e}^{\Del} _{\perp,1} \cdot (\vec{x}_{\Del}-\vec{x}_{\Del'}) | +
 2^m |\vec{e}^{\Del}_{\perp,2} \cdot (\vec{x}_{\Del}-\vec{x}_{\Del'}) | + 2^{m-j/2} |x_{\Del}^0-x_{\Del'}^0| \right) \nonumber \EEA
while $d_{ang}(\Del,\Del')\approx 2^{j/2} |\vec{l}|$. 
Note the supplementary scaling factor $|\vec{l}|$ in the previous expression, a consequence of the 'frustration' due to the non-alignment of $e_{\Del}$ and $e_{\Del'}$. 
However this supplementary factor is not needed for the convergence and may be simply forgotten. One reduces to  a problem similar to  (\ref{eq:Z4}) but in $\Z^6$, where the two supplementary dimensions
account for the angle coordinates $l_1,l_2$. 

\item[(iii)] (general case) Assume e.g. that $m<m'$ and $\vec{e}_{\Del}=(1,0,0)$, say.  Then the transversal region $x_2\in [(k_2-1) 2^{-m},(k_2+1)2^{-m}]$ (corresponding to one of the
transversal sides of  $\Del$) contains $O(2^{m'-m})$ integer multiples of $2^{-m'}$, corresponding to different spatial localizations for $\Del'$. Summing over all of these
cost a factor $O(2^{|m'-m|})$. The same counting argument goes for the four space-time directions and the angular direction. All together one loses a volume factor which is at most
$$\approx 2^{2|(j/2-j'/2)|+|(m+j/2)-(m'+j'/2)|+2|m-m'|+|(m-j/2)-(m'-j'/2)|})\approx 2^{4|m-m'|+2|j-j'|},$$ exactly the  same factor found in eq. (\ref{eq:almost-ortho}). Consider
a perpendicular direction $x_{\perp,1}$, say, and assume e.g. that $m\le m''$. Summing over all
intermediary scales $m<m'<m''$ and measuring $x_{\perp,1}$ is terms of the intermediary scaling $2^{-m'}$ yields a volume factor $2^{|m-m'|}$  for the sum of
$d(\Del,\Del')$  over $\Del'$. On the other hand, the scaled distance $d(\Del,\Del'')$ should be measured in terms of the lower scaling $2^{-m}$ and not $2^{-m'}$, which yields the same
volume factor in the right-hand side of  (\ref{eq:decay-summability}). Without the  prefactors in $1+|m-m'|^2, 1+|m'-m''|^2, 1+|m-m''|^2$ in eq. (\ref{eq:almost-ortho}), one would lose only a logarithmic factor
$O(|m-m''|)$ to pay for the sum over all intermediary scales. A simple exponential factor $2^{-\alpha|x-x''|}$ would be useless since $2^{-\alpha|x-x'|}2^{-\alpha|x-x''|}=
2^{-\alpha|x-x''|}$ and the logarithmic factor would remain; but $\sum_{m'=m}^{m''} (1+|m-m'|^2)^{-r} (1+|m'-m''|^2)^{-r}=O(1)$ provided $r>1$. 
Keeping the assumption $m\le m''$, one must still sum over $m'<m$ and $m'>m''$. If $m'<m$ then (measuring $x_2$ in terms of the intermediate scale $2^{-m}$) a volume factor $O(2^{|m-m'|})$ is lost  on the left-hand side of (\ref{eq:decay-summability}), and nothing on the right-hand side. The scaling factors make up for this loss since 
\BEA &&  \sum_{m'\le m} 2^{|m-m'|}\cdot \left(\frac{2^{-|m-m'|}}{1+|m-m'|^2}\right)^r \left(\frac{2^{-|m'-m''|}}{(1+|m'-m''|^2)}\right)^r \nonumber\\
&& \le C \left(\frac{2^{-|m-m''|}}{1+|m-m''|^2}\right)^r
\sum_{m'\le m} 2^{-(2r-1)|m-m'|} \nonumber\\ && \le C'\left(\frac{2^{-|m-m''|}}{1+|m-m''|^2}\right)^r  \label{eq:scaling-make-up} \EEA
provided $2r-1>0$. If on the contrary $m'>m''$ then (measuring $x_2$ in terms of the intermediate scale $2^{-m''}$) a volume factor $O(2^{|m-m'|})$, resp. $O(2^{|m-m''|})$  is lost  on the left-hand side, resp. on the right-hand side of (\ref{eq:decay-summability}), all together one has lost a factor $O(2^{|m'-m''|})$ which is compensated by the scaling factors exactly as in eq. (\ref{eq:scaling-make-up}).

\end{itemize}
   \hfill \eop

%%%%%%%%%%%%%%%%%%%%%%%%%%%%%%%

\section{Ultra-violet bounds for Klein-Gordon operator on Minkowski space-time}

We consider here the Klein-Gordon operator on flat space-time, $\Box_{\mu}=\partial_t^2-\sum_{i=1}^3 \partial_{x_i}^2 -\mu^2$ $(\mu\ge 0)$. General bounds such as the one given in the
Main Theorem are greatly improved by the use of Minkowski curvelets which are adapted to $\Box_{\mu}$ rather than to the wave operator $\Box_0$. Because $\Box_{\mu}$ has constant coefficients
and is so simple, the Klein-Gordon curvelets $(\psi_{\Del})$ we shall be using here are a straightforward modification of those used previously. Up to a rescaling, one may assume that
$\mu=1$; however, in order to keep track of the $\mu$-dependence, we shall simply assume that $\mu\approx 1$, which allows us to keep the coarse scale at $j=0$. One then simply
substitutes $\xi^2-\mu^2$ to $\xi^2$ in eq. (\ref{eq:LMM}), so that (\ref{eq:1.13}) becomes
\BEQ |\xi^2|\approx  2^{2m}, \qquad ||\xi_0|-E(\vec{\xi})|\approx 2^{m-j/2}, \qquad |\xi_0|\approx E(\vec{\xi})\approx 2^{m+j/2} \label{eq:1.13bis} \EEQ 
with $E(\vec{\xi}):=\sqrt{|\vec{\xi}|^2+\mu^2}$.

Lemmas 
\ref{lem:mol}, \ref{lem:almost-ortho}, \ref{lem:decay-sum} still hold provided one redefines $\nabla_{//}$, $e_{\Del}$ and $e_{\Del}^*$ as follows. Let $\xi_{\Del}$ be the 'center' of the
support of $\hat{\psi}_{\Del}$ (any point in $\supp\, \hat{\psi}_{\Del}$ would do equally well). The derivative operator $\nabla_{//}$ is defined to be parallel to the hyperboloid
$\xi^2-\mu^2=0$ at $\xi_{\Del}$, so that $\nabla_{//}(\xi^2-\mu^2)\big|_{\xi_{\Del}}=0$, and contained in the propagation plane. This fixes $\nabla_{//}$ up to normalization. We choose 
$\nabla_{//}:=\partial_{\xi_0}+\frac{\xi_0^{\Del}}{\xi_1^{\Del}} \partial_{\xi_1}$ if $\vec{e}_{\Del}=(1,0,0)$. Then $e_{\Del}=(1,\frac{\xi_0^{\Del}}{\xi_1^{\Del}},0,0)$ and
$e_{\Del}^*:=(\frac{\xi_0^{\Del}}{\xi_1^{\Del}},1,0,0)$ is chosen to be Minkowski orthogonal to $e_{\Del}$. 

Recall \cite{BogShi} that the retarded Green function of the Klein-Gordon operator is given in Fourier coordinates by $\hat{G}(\xi)=\lim_{\eps\to 0} \frac{1}{\xi^2-\mu^2-\II \eps\sgn(\xi_0)}$. 
The causal Green function $G^c$ is given instead by $\lim_{\eps\to 0} \frac{1}{\xi^2-\mu^2-\II \eps}$, and other Green functions are linear combinations of either $G$ or $G^c$ with
${\bf 1}_{\xi_0>0} \del(\xi^2-\mu^2)$ and ${\bf 1}_{\xi_0<0} \del(\xi^2-\mu^2)$. Since Klein-Gordon curvelets are supported outside the hyperboloid, the following lemma holds equivalently
for any of these Green functions.

We now prove the main theorem, which we restate for the convenience of the reader. Its proof follows along the same lines as the previous estimates.

\begin{Theorem}[off-diagonal decay of Green function]

Let $\Del\in\D^{m,j},\Del'\in\D^{m',j'}$, with $m,m'\in\Z$, $j,j'\ge 1$, $m+j/2,m'+j'/2\ge 0$. Then, for every $N\ge 0$, there exists a constant $C_N>0$ such that

\BEQ |G_{\Del,\Del'}|\lesssim 2^{-2\sup(m,m')} d(\Del,\Del')^{-N}.\EEQ

\end{Theorem}

{\bf Proof.}

 By space-time translation invariance, we may assume that $\Del'=(m',j',\vec{e}',(0,0,0,0))$ is
centered at the origin of space-time and use the improved Fourier decay (\ref{eq:Fourier-decay-centered}). Since $\left|\frac{1}{\xi^2-\mu^2}\right|\lesssim 2^{-2\max(m,m')}$ for
$\xi\in\supp(\hat{\psi}_{\Del})\cap \supp(\hat{\psi}_{\Del'})$, the Fourier part of the decay, i.e. the bound $|G_{\Del,\Del'}|\lesssim 2^{-2\max(m,m')} (d_{\xi}(\Del,\Del'))^{-N}$, follows
as in (i). We shall also prove the space-decay in Fourier space this time by using a Fourier transformation, assuming for simplicity that the Fourier indices 
$(m,j,\vec{e}_{\Del})$ and $(m',j',\vec{e}_{\Del'})$ coincide, so that  $\psi_{\Del}$ is obtained from $\psi_{\Del'}$ by a translation in real-space. By definition of $G$ and of the Minkowski curvelets,
\BEQ G_{\Del,\Del'}=\int \frac{d\xi}{\xi^2-\mu^2} |\psi_{\Del'}(\xi)|^2 u_k(\xi),\EEQ
where 
\BEQ u_k(\xi)=\exp \II 2^{-m} \left\{ (k_0 2^{j/2}+k_1 2^{-j/2})\xi_0 -(k_0 2^{j/2}-k_1 2^{-j/2})\xi_1-\vec{k}_{\perp}\cdot\vec{\xi}_{\perp} \right\}\EEQ
is an oscillation proportional to $d_x(\Del,\Del')$. Then we rewrite $u_k(\xi)$ as $\frac{\nabla_{//}u_k(\xi)}{2^{-m-j/2}k_1}$, resp. $\frac{\nabla_{\perp}u_k(\xi)}{2^{-m}k_{\perp}}$
(along the two transversal directions) or $\frac{\nabla_{off}u_k(\xi)}{2^{-m+j/2}k_0}$, and use an integration by parts. Bounds for $\nabla\hat{\psi}_{\Del}(\xi)$,
$\nabla=\nabla_{//},\nabla_{\perp}$ or $\nabla_{off}$, follow from eq. (\ref{eq:Fourier-decay-centered}), yielding precisely the inverse of the above scaling factors, $2^{-m-j/2}$, resp.
$2^{-m},2^{-m+j/2}$. Similarly
\BEA && \left|\nabla_{//} \left(\frac{1}{\xi^2-\mu^2}\right) \right| \lesssim \frac{2^{-m-j/2}}{\xi^2-\mu^2}, \quad
\left|\nabla_{\perp} \left(\frac{1}{\xi^2-\mu^2}\right) \right| \lesssim \frac{2^{-m}}{\xi^2-\mu^2},\nonumber\\ &&\qquad \qquad \qquad \qquad
  \left|\nabla_{off} \left(\frac{1}{\xi^2-\mu^2}\right) \right| \lesssim \frac{2^{-m+j/2}}{\xi^2-\mu^2}.\EEA
  For the bound concerning $\nabla_{//}$ we used the following fact (written for $\vec{e}_{\Del}$ along the first axis of coordinates): 
\BEQ \half\nabla_{//}(\xi^2-\mu^2)=\xi_0-\frac{\xi_0^{\Del}}{\xi_1^{\Del}}\xi_1=(\xi_0-\xi_0^{\Del})+\frac{\xi_0^{\Del}}{\xi_1^{\Del}}(\xi_1^{\Del}-\xi_1)\lesssim 2^{m-j/2}.\EEQ

These arguments generalize to curvelets $\Del,\Del'$ with different Fourier indices by translating the curvelet whose window has a  larger side. \hfill\eop

%

%%%%%%%%%%%%%%%%%%%%%%%%%%%%%%%%%%%%%%%%%%%
%%%%%%%%%%%%%%%%%%%%%%%%%%%%%%%%%%%%%%%%%%%

%%%%%%%%%%%%%%%%%%%%%%%%%%%%%%%%%%%%%%%%%%%%%%%%%
%%%%%%%%%%%%%%%%%%%%%%%%%%%%%%%%%%
%%%%%%%%%%%%%%%%%%%%%%%%%%%%%שש
%%%%%%%%%%%%%%%%%%%%%%%%%%%%%%
%%%%%%%%%%%%%%%%%%%%%%%%%%

\end{document}